\documentclass[onecolumn,preprintnumbers, amsmath,amssymb,nopreprintnumbers,nopacs]{revtex4}
\usepackage[colorlinks,urlcolor=blue]{hyperref}
\usepackage{graphicx}%Include figure files
\usepackage{lineno}
\usepackage{dcolumn}% Align table columns on decimal point
\usepackage{bm}% bold math
\usepackage{appendix}
\usepackage{verbatim}

\usepackage{amsmath}
\usepackage{enumerate}
\usepackage{soul,xcolor}

\makeatletter
  \let\@font@info\@gobble
  \let\@font@warning\@gobble
\makeatother

\begin{document}
%\linenumbers
%\preprint{APS/123-QED}

\title{Large deviations for continuous time random walks}

 %\altaffiliation[Also at ]{School of Mathematics and Statistics, Lanzhou University.}%Lines break automatically or can be forced with \\

\author{Wanli Wang}%
\affiliation{%
Department of Physics, Institute of Nanotechnology and Advanced Materials, Bar-Ilan University, Ramat-Gan
52900, Israel}

\author{Eli Barkai   }%
\affiliation{%
    Department of Physics, Institute of Nanotechnology and Advanced Materials, Bar-Ilan University, Ramat-Gan 52900, Israel}
\author{Stanislav Burov}%
\affiliation{%
    Department of Physics, Bar-Ilan University, Ramat-Gan 52900, Israel}

%Department of Physics, Institute of Nanotechnology and Advanced Materials, Bar-Ilan University, Ramat-Gan
%52900, Israel
%\\
%$^3$Physics Department T30g, Technical University of Munich, 85747 Garching, Germany
 %}%

% \homepage{http://www. Second. institution. edu/~Charlie. %Author}
%\affiliation{

\date{\today}% It is always \today,  today,

\begin{abstract}
Recently observation of random walks in complex environments like the cell and other  glassy systems
revealed that the spreading of particles, at its tails, follows a spatial  exponential decay instead of the canonical Gaussian. We use the
widely applicable continuous time random walk model and obtain the large deviation description
of the propagator. Under mild conditions that the microscopic
 jump lengths distribution is decaying exponentially or  faster
i.e. L\'evy like power law distributed  jump lengths are excluded,
and that the distribution of the waiting times is analytical for short
waiting times, the spreading of particles follows an exponential decay at large
distances, with a logarithmic correction. Here we show how anti-bunching of jump events reduces the effect, while bunching and intermittency enhances it. We employ  exact solutions of the continuous time random walk model to test the large deviation theory.
%
%
%
%the low order moments of $F, B$ and $Z$ are governed by the density of the typical fluctuations and high order of moments are determined by the rare events.

%high order of the absolute moment  are

%deviation case but not for %$\varepsilon=n-t/\langle\tau\rangle$

% Aging continue time random walk has been of much interesting as it is one of the fundamental processes in many nonlinear systems. In this paper, we consider functionals of anomalous diffusion, the  forward and backward  Feynman-Kac for functionals of anomalous paths are derived based on aging continue time random walk, Furthermore, both  power law  and tempered power law waiting time are considered. Especially, Setting $t_a=0$, which is agree with precious results. Finally,  we give a application of the Feynman-Kac equation.
\end{abstract}

%\pacs{02. 50. -r,  05. 20. -y,  05. 40. -a }% PACS,  the Physics and Astronomy
% url
%https://publishing.aip.org/publishing/pacs/pacs-reg00#05
%
%   02. 50. -r   Probability theory, stochastic processes, and statistics
%  05.30.Pr	Fractional statistics systems (anyons, etc.)

% 05.40.-a Fluctuation phenomena, random processes, noise, and Brownian motion (for fluctuations in superconductivity, see 74.40.-n; for statistical theory and fluctuations in nuclear reactions, see 24.60.-k; for fluctuations in plasma, see 52.25.Gj; for nonlinear dynamics and chaos, see 05.45.-a)
%
%��05.10.Gg Stochastic analysis methods (Fokker-Planck, Langevin, etc.)

% 02.50.Ey	Stochastic processes
%  45.10.Hj	Perturbation and fractional calculus methods

%\keywords{Suggested keywords}%Use showkeys class option if %keyword
\maketitle

\section{Introduction}

Following the erratic motion of pollen under his microscope Robert
 Brown discovered
what is called today: Brownian motion. This phenomenon was  modeled by Einstein
and others, with random
walk theory  while the  mathematical description of Brownian motion, i.e. the Wiener
process \cite{Majumdar2005Brownian}, was quickly established soon after. Two ingredients of this widely
observed dynamics are that the mean square displacement increases linearly with
time and that   the spreading of the packet of particles all starting
on a common origin is Gaussian \cite{Montroll1965Random,Haus1987Diffusion}. The latter is the physical manifestation
of the central limit theorem. However, in Reference \cite{Pinaki2007Universal}  Chaudhuri,  Berthier, and Kob, discovered what might turn
out to be a no less universal feature of random walks. Analysing experimental
and numerical data, e.g. the motion of particles in glass forming systems,
they revealed an exponential decay of the packet of spreading particles; see related works in References \cite{Kege2000Direct,Masolivera200dynamic,Weeks2000Three,Pinaki2007Universal,Wang2009Anomalous,Hapca2009Anomalous,Leptos2009Dynamics,Eisenmann2010Shear,Toyota2011Non,Skaug2013Intermittent,Xue2016Probing,Wang2017Three,Jeanneret2016Entrainment,Chechkin2017Brownian,Cherstvy2019Non,Witzel2019Heterogeneities,Shin2019Anomalous,Singh2020Non,Mejia2020Tracer,Xue2020Diffusion}.
Soon after this, similar behavior described by the Laplace distribution, was found for many
other systems, including for molecules in the cell environment \cite{Weeks2000Three,Wang2012Brownian,Munder2016transition}.
In \cite{Pinaki2007Universal} it was suggested that a specific type of continuous time random
walk (CTRW) model  could explain the physics of the observed behavior. Two of us have
recently shown that the phenomenon is indeed universal as it holds in general \cite{Barkai2020Packets}
and under very mild conditions (see below). The goal of this
paper is to produce further evidence for the phenomenon, present exact
solutions of the model and compare it with the new theory. We  also present
a more
detailed analysis of the model covering cases not discussed previously.

The observed behavior, is related to the way
experimentalists record the erratic paths. Following many
trajectories one may find universal features
of the motion in at least two different ways.
In some situations the measurement time is very long such that many jumps occurred and consequently Gaussian statistics will take place.
However, in practice  experiments  are not conducted
for an infinite time.
In fact in many experiments, the motion follows some trapping and then released motion,
i.e. a hop and then wait dynamics, as modeled by the CTRW.
In these experiments one follows many paths, e.g. many single molecules,
however typically each trajectory is recorded separately.
When    the averaged  number of jumps recorded
under the microscope (or in simulations) within the observation time
$(0,t)$  is not large, one would expect
naively that  this
would imply the non-existence of universal  statistical laws. However a second limit
is important, which is large positions (and  finite time). In this case,
we will promote the idea  that exponential
spreading is the rule.
Thus the observation of either  Gaussian or exponential spreading depends on the time scale
and length scale of the motion and in statistical sense the  temporal and spatial extent of the field
of view.  The idea is that by observing the density of packet of particles in its tails (large $x$) we are in fact  considering trajectories where the number of jumps is large, compared to the mean. We advocate that universal laws,  that demand a large number of steps, can be found for large displacements (and fixed time) or for large time as usual.
Theoretically since we have two parameters that can be made large,
i.e. $t/\langle \tau \rangle$ or $x/\sigma$ we may obtain more than
one limiting law.
Here $\langle \tau\rangle$ is the average time between jumps,
and $\sigma$ is the variance of jumps lengths.
The motion is unbiased
hence mean jump size is zero.
We  will consider, among other things the case, when both $t/\langle \tau \rangle$ and $x/\sigma$ are
large  and their ratio is finite.

The remainder of the manuscript is organized as follows.  In Sects. \ref{shahshhawaa} and \ref{sjdhhsiwq}, after presenting the CTRW model, we consider the far tail of the distribution of the number of renewals and the position of the random walker, where the waiting times are drawn from the exponential and the Erlang distributions, respectively. The bunching effect is investigated in Sect.~\ref{sahshawa} using the sum of two exponential probability density functions (PDF).
Finally, we conclude  with a discussion.

\section{Appetizer for exponential tails}\label{shahshhawaa}
We consider the well known CTRW model \cite{Montroll1965Random,Kindermann2017Nonergodic,Kutner2017continuous}.
This model describes a wait and then jump process. A particle
in dimension one, starts on the origin at time $t=0$. We draw a waiting
time from the  PDF $\psi(\tau)$ and then after the wait, the particle
will perform a jump whose length, $\chi$, is drawn from $f(\chi)$. The process is
then renewed. In our case the waiting times and the jump lengths are not correlated.
 The  position of the particle at time $t$
is $x(t) = \sum_{i=1} ^n \chi_i$, and here $n$ is the random number of jumps
within the time of observation $(0,t)$. The focus of this
manuscript is the  PDF of finding the particle
on $x$ at time $t$ which is denoted  $P(x,t)$. In particular the large $x$ limit
of this distribution is of interest. We focus on non-biased CTRWs,
and then if the mean waiting times and the variance of jump lengths are finite the density $P(x,t)$ will converge to a Gaussian as
expected from the central limit theorem. This limit theorem is valid for $t\to \infty$ and $x \propto  t^{1/2}$, while here we are interested in finite time
effects and the large $x$ limit, to be defined more precisely below.
Generally, the solution of the model is given by
\begin{equation}\label{PXT101}
P(x,t) = \sum_{n=0} ^\infty  Q_t (n) P(x|n).
\end{equation}
Here the sum is over the possible outcomes of the number of jumps
in the process, $Q_t(n)$ is the probability of attaining $n$
jumps \cite{Godreche2001Statistics}, while $P(x|n)$ is the probability of finding the particle
on $x$ conditioned it made $n$ jumps. In Laplace space~\cite{Barkai2020Packets},
\begin{equation}\label{LaplaceQSN}
\hat{Q}_s (n) =\int_0^\infty Q_t(n)\exp(-st)dt= \frac{1 - \hat{\psi} (s)}{ s} \hat{\psi}^n (s),
\end{equation}
where $s$ is the Laplace pair of $t$.
In particular, when $n=0$, $Q_t(n=0)=\int_t^\infty \psi(\tau)d\tau$ is called the survival probability.

Generally finding $Q_t(n)$ and $P(x|n)$ is non trivial. Here the goal is to
consider special choices of jump length distributions $f(\chi)$ and waiting
time PDFs  $\psi(\tau)$ which allow us to express the solution as
an infinite sum over $n$ explicitly, i.e. without restoring to inverse
Fourier and Laplace transforms and/or numerical simulations.
This allows us to compare between
exact solutions of the problem, and the large
deviation  theory we have promoted previously \cite{Barkai2020Packets}. Further, with the specific choices
of the input distributions $\psi(\tau)$ and $f(\chi)$ we can derive
large deviations theory in a straight forward way.
The price that we pay is that we do not consider here the theory in its full
generality, since we stick to exactly solvable models.

\subsection{Displacement follows Gaussian distribution}
We start the analysis with the simplest choice of waiting times and jump length
distributions. The jump length PDF
 is assumed to be Gaussian
with zero mean
%\added{Should we say 'mean zero'? see words below 'and' }
and variance $\sigma=1$, i.e., $f(\chi) = \exp( - \chi^2/2)/\sqrt{2 \pi}$. It then follows
that $P(x|n)$ is  Gaussian as well.
The waiting times are exponentially distributed
$\psi(\tau) = \exp(- \tau)$ for $\tau>0$. Hence the mean waiting time
is $\langle \tau \rangle =1$ and $Q_t (n)$ obeys
Poisson statistics.   The density of spreading particles all starting on the origin is
therefore
\begin{equation}
\begin{split}
P(x,t)= &\sum_{n=0} ^\infty \frac{ t^n \exp( -t) }{n!} \frac{\exp( - x^2 / 2 n)}{ \sqrt{ 2 \pi n} }\\
\end{split}
\label{eq01}
\end{equation}
and in this sense we have an exactly solvable model valid for any time
and $x$. Here the term $n=0$ contains a delta function on the origin, which of course is not plotted  in the figures we present below. This describes particles not moving at all.

%Here the second line in %the right hand sided of %Equation~\eqref{eq01} is %related to non-moving %particles.

To analyse this sum in the large $x$ limit we use  Cram\'{e}r-Daniels approach \cite{Daniels1954Saddlepoint} to large
deviations. Essentially this is a saddle point method, exploiting the fact that
one may relate the cumulant generating function of $P(x,t)$ and the large deviations
in the system. For that let us take the Fourier transform of Equation
\eqref{eq01}
and then we have
\begin{equation}
\widetilde{P}(k,t)= \sum_{n=0}^\infty { \frac{t^n \exp(-t)}{n!}} \widetilde{f}^n (k),
\label{eq02}
\end{equation}
where we used the Fourier transform of $f(\chi)$ denoted $\widetilde{f}(k)=\exp( - k^2 /2)$.
We have exploited the fact that the jump lengths are independent and identically (IID) hence from
convolution theorem of Fourier transforms  we have the expression
$\widetilde{P}(k|n)=\widetilde{f}^n(k)$.
The moment generating function is $\langle \exp( u x) \rangle= \int_{-\infty} ^\infty P(x,t) \exp( u x) {\rm d} x$.
Replacing $k \to - i u$ and summing the series
\begin{equation}
\langle \exp( u x) \rangle = \exp\left[ - t(1 - e^{u^2/2} ) \right].
\end{equation}
Then the cumulant generating function is simply the log of the above expression, we
denote it
\begin{equation}
K(u) = \ln \left[ \langle \exp( u x) \rangle\right] = - t \left( 1 - e^{u^2/2} \right).
\end{equation}
Now we are interested in the large $x$ limit of $P(x,t)$, which by definition
is  the inverse Fourier transform $\widetilde{P}(k,t)$. Instead of integrating over
$k$ one may switch to integration over $u$ and using saddle point
method valid for the large $x$
one then finds a well known large deviation formula \cite{Daniels1954Saddlepoint}
\begin{equation}\label{Largedeviation101}
P(x,t) \sim {1 \over \sqrt{ 2 \pi K''(\hat{u} ) } } \exp\left[ K\left( \hat{u} \right) - \hat{u} x \right].
\end{equation}
Here $\hat{u}$ is given by the solution $K'(u) = x$. While in this section we are treating a special
choice of jump length and waiting time, this tool will serve us all along the paper.
One should notice that a cumulant generating function does not always exist. For example
if $f(\chi)$ decays as a power law for large $x$, the moments
of the process will diverge. From here we see the first condition
of the theory to hold in generality:  we assume that cumulant generating function
of $f(\chi)$ and $P(x,t)$ exist, so the decay of $f(\chi)$ is faster than exponential
\cite{Eli2020SRW}.

\begin{figure}[htb]
 \centering
 \includegraphics[width=0.5\textwidth]{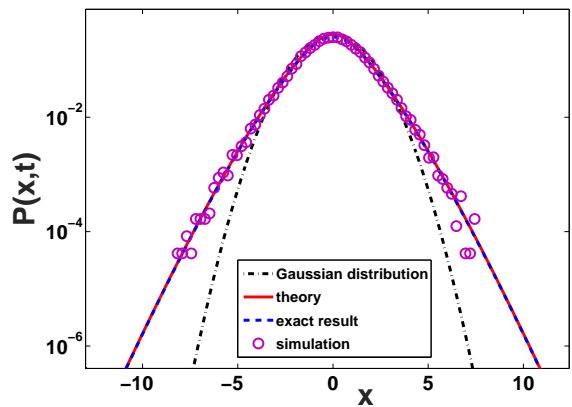}\\
 \caption{The distribution $P(x,t)$ with exponentially  distributed  waiting times and Gaussian displacements. The time is $t=2$ and for simulations we used $5\times10^7$ trajectories. Our theory Equation~\eqref{eqWWW} performs perfectly, while  the Gaussian distribution Equation~\eqref{ExpGaussian}, black dash-dotted line, completely fails for the far tails of the distribution of the position. Note that the theoretical prediction Equation~\eqref{eqWWW} works extremely well also for the central part of the distribution. Here the exact result is obtained from Equation~\eqref{eq01}.
}\label{ExpGaussianPXT}
\end{figure}

 For the Gaussian choice of $f(\chi)$, to find $\hat{u}$ we have
\begin{equation}
K'(\hat{u}) = t \hat{u} \exp\left( { \hat{u}^2 \over 2}\right) = x.
\end{equation}
The solution of this equation is given by the Lambert $W_0$ function \cite{Dence2013brief} where the subscript zero
denotes the branch of this function. The latter is the solution of the equation $y \exp(y) = z$ given
by $y=W_0(z)$ if $z \ge 0$. Hence we get
\begin{equation}\label{ussss1}
\hat{u} = \pm\sqrt{ W_0 \left[ \left({ x \over t} \right)^2  \right]},
\end{equation}
where  the sign of $\hat{u}$ is dictated by the sign of $x$.
Using Equation~\eqref{Largedeviation101} we then find
\begin{equation}
P(x,t) \sim  {\exp\left( -  t - |x|\left( \sqrt{ W_0 [ \frac{x^2}{t^2}] }  - {1 \over \sqrt{ W_0 [ \frac{x^2}{t^2}]}}\right) \right)
\over \sqrt{ 2 \pi
K''(\hat{u})} }
\label{eqWWW}
\end{equation}
with
$$K''(\hat{u}) = |x| \left( \sqrt{W_0[(x/t)^2]} + \frac{1} {\sqrt{ W_0[(x/t)^2]}}\right).$$
This is plotted in Figure ~\ref{ExpGaussianPXT}.
Using the large and small $z$ limits of the Lambert function, i.e. $W_0(z^2) \sim 2 \ln(z)$ when $z>>1$
and $W_0 (z^2)\sim z^2 $ for $z<<1$, we find the following two  limits:
\begin{equation}
P(x,t) \simeq \exp \left[ - x \sqrt{ 2 \ln(x/t) } \right]
\end{equation}
valid when $x/t >> 1$, and
\begin{equation}\label{ExpGaussian}
P(x,t) \simeq \exp[  -x^2 / 2 t]
\end{equation}
in the opposite limit when $|x|/t\to 0$.
The first is the mentioned exponential decay, and it includes what we call
 the  $\ln(x)$ correction.  As far as we know, in the experimental literature this $\ln(x)$
is not reported. However this fact is not surprising as the $\ln(x)$ is clearly a slowly
varying function and hence difficult to detect in reality.
We note that the $\log$ correction implies that the decay is in reality slightly faster than exponential, implying the existence of the cumulant generating function.
 The second case is the well known  Gaussian behavior found in the
centre of the packet.

\begin{figure}[htb]
 \centering
 \includegraphics[width=0.5\textwidth]{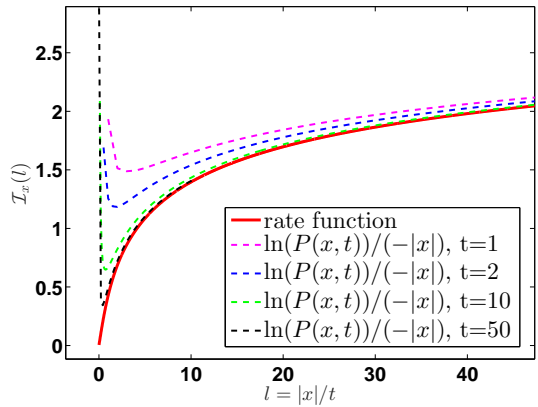}\\
 \caption{The plot of the rate function $\mathcal{I}_x(l)$ with $l=|x|/t$. The red solid line, predicted by Equation~\eqref{ratePosi}, describes the rate function $\mathcal{I}_x(x/t)$ capturing the behavior of large deviations. Note that for large $l$ we have $\mathcal{I}_x(l)\sim \sqrt{2\ln(l)}$ and this means that $P(x,t)$ is decaying exponentially with the distance $x$ with a  correction.
 The dashed lines are the plot of $\ln(P(x,t))/(-|x|)$ calculated from  Equation~\eqref{eq01}.
}\label{RateFunctionVersusX}
\end{figure}

\begin{figure}[htb]
 \centering
 \includegraphics[width=0.5\textwidth]{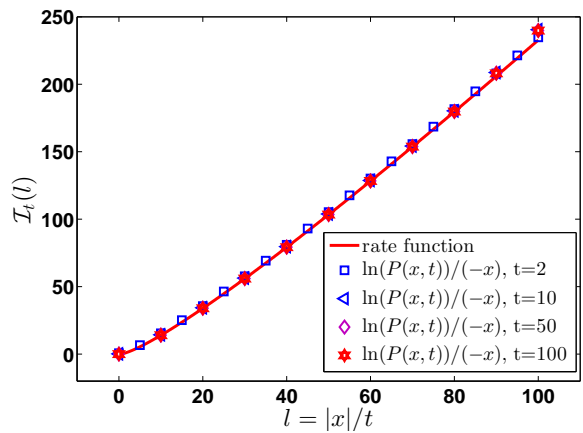}\\
 \caption{Rate function $\mathcal{I}_t(l)$  versus $l=|x|/t$ for different $t$. The red solid line is the (time) rate function Equation~\eqref{eXPRateTcorrect}. As in Figures \ref{ExpGaussianPXT} and \ref{RateFunctionVersusX} here $\psi(\tau)=\exp(-\tau)$ and $f(\chi) =\exp(-\chi^2/2)/\sqrt{2 \pi}$.
}\label{RateFunctionVersusT}
\end{figure}

We may further  formulate our results using two very much
 related approaches,
both consider  the limits
$x\to \infty$ and $t \to \infty$, and their ratio
$x/t = l$ is kept fixed. Using Equation~\eqref{eqWWW} the first limiting law is
\begin{equation}\label{ratePosi}
\lim_{x\to \infty} -\ln( P(x,t)) /x = \mathcal{I}_x(l)
\end{equation}
with
$\mathcal{I}_x (l) = l^{-1} - 1/\sqrt{ W_0(l^2))} + \sqrt{ W_0(l^2)} $
and the second refers to
\begin{equation}\label{eXPRateTcorrect}
\lim_{t\to \infty} -\ln( P(x,t)) /t = \mathcal{I}_t(l)
\end{equation}
with $\mathcal{I}_t(l) =l \mathcal{I}_x(l)$.
The functions $\mathcal{I}_t(l)$ and $\mathcal{I}_x(l)$ are called the rate functions \cite{Touchette2009large,Daniel2018Anomalous,Touchette2018Introduction}.
All these approaches are essentially identical.  Most experimentalists
in the field plot $P(x,t)$  versus $x$ on a semi-log scale, and for that aim
Equation (\ref{eqWWW}) including the prefactor $1/\sqrt{ 2 \pi K''(\hat{u})}$
is useful. Mathematicians  influenced by long time  (large $x$)  ideas,
and the large deviation literature,  would
possibly find it natural to use $\mathcal{I}_t(l)$ ($\mathcal{I}_x(t)$) respectively.
In demonstration below we plot the exact solution for finite times,
using both approaches; see Figures \ref{ExpGaussianPXT}, \ref{RateFunctionVersusX},  and \ref{RateFunctionVersusT}.

 We now present a simple explanation for the exponential tail using a lower bound. Clearly
$$P(x,t)\geq {\exp(-\frac{x^2}{2n^*})\over\sqrt{2\pi n^*}}{t^{n^*}\exp(-t)\over(n^*)!}$$
and here $n^*$ is the value of $n$ for which the summand in Equation~\eqref{eq01} is the maximum. To find $n^*$, we use Stirling's approximation, i.e., $n!\sim \sqrt{2\pi n}(n/e)^n$, and
\begin{equation}
\frac{d}{dn}\exp\left(-t+n\ln(t)-\frac{x^2}{2n}-\ln\left(\sqrt{2\pi n}\left(\frac{n}{e}\right)^n\right)\right)=0.
\end{equation}
The solution of the above equation is $n^*\sim |x|/\sqrt{W_0(x^2/t^2)}$.
As expected, in the limit $|x|/t\to \infty$, we find the exponential decay again
%$$P(x,t)\geq \exp\left(-t+n^*\ln(t)-\frac{x^2}{2n^*}-n^*\ln(n^*)+n^*\right).$$
\begin{equation}
P(x,t)\geq \exp\left(-t-|x|\sqrt{\frac{\ln(\frac{|x|}{t})}{2}}+|x|\frac{\ln(\frac{te}{|x|}\sqrt{2\ln(\frac{|x|}{t}  })}{\sqrt{2\ln(|x|/t)}}\right),
\end{equation}
where we used the  asymptotic behavior of  the Lambert  function,  i.e., $W_0(|x|)\sim \ln(|x|)$ with $|x|\to\infty$.
The claim in \cite{Barkai2020Packets} is much more general as similar behavior is found for a large class of waiting times  and jump lengths PDFs, see details below.

%{\reform This is not main text}
%Since we have the term $(n^*)!$
%\begin{equation}
%\begin{split}
%\frac{1}{(n^*)!}&\sim \frac{1}{\sqrt{2\pi n^*}(n^*/e)^{n^*}}\\
%&\sim \exp(n^*\ln(e/n^*))\\
%&=\exp(\frac{|x|}{W_0(x^2/t^2)}\ln(\frac{eW_0(x^2/t^2)}{|x|}))
%\end{split}
%\end{equation}
%{\reform We should delete this}

%\begin{figure}[htb]
% \centering
 % Requires %\usepackage{graphicx}
% \includegraphics[width=0.5\textwidth]{RateFunctionVersusXV2.eps}\\
% \caption{\added{Same as Figure ~\ref{RateFunctionVersusX}. Here the symbols are the plot of $\ln(P(x,t)\sqrt{ 2 \pi
%K''(\hat{u})})/(-|x|)$ showing perfect match.}
%}\label{RateFunctionVersusXV2}
%\end{figure}

\subsection{Displacement drawn from the discrete PDF}
We now investigate a second example, which further demonstrates the exponential
far tail of $P(x,t)$. Consider a CTRW on a one dimensional lattice, with exponential waiting times. Hence $f(\chi) = [ \delta(\chi-1) + \delta(\chi+1)]/2$
and $\psi(\tau) = \exp(-\tau)$. The Fourier transform of the jump length distribution is $\widetilde{f}(k) = \cos(k)$ and the moment generating function
is $\langle \exp( u \chi)\rangle = \cosh(u)$, where $\cosh(u)$  is hyperbolic  cosine function. Here $P(x|n)$ is the well
known Binomial distribution
\begin{equation}\label{eqes16}
P(x|n) = {1 \over 2^{n+1}}[1+(-1)^{n+x}]{ n! \over \left(\frac{n+x}{2}\right)! \left(\frac{n-x}{2}\right)!},~~~~x\leq n,
\end{equation}
where  $x$ is an integer since the random walker is on a lattice with a unit spacing.
Notice that when $n$ is odd/even the same holds for $x$. The CTRW probability is then
\begin{equation}
P(x,t) = \sum_{n=0} ^\infty { e^{- t} t^n \over n!} P(x|n).
\end{equation}
Switching to  the Fourier space from convolution theorem, the Fourier transform
of $P(x|n)$ is $\widetilde{f}^n(k)$.
Then
after summation it is easy to find the moment generating function
$M(u)=\langle \exp(u x)\rangle = \exp[ - t ( 1 - \cosh(u))]$. We then use,
as before, the cumulant generating function $K(u) = - t [ 1 - \cosh(u) ]$.
We now need to find the solution of $ K^{'}(u) = x$ which is
given by $\hat{u}=\sinh^{-1} (x/t)=\ln(x/t+\sqrt{1+x^2/t^2})$, then using Equation~\eqref{Largedeviation101}
we get
\begin{equation}
P(x,t) \sim \frac{\exp\left[ - x \sinh^{-1}\left({x\over t}\right) - t + t \sqrt{ 1 + { x^2 \over t^2} } \right]}{\sqrt{2\pi K^{''}(\hat{u})}}
\label{pxtexpdelta}
\end{equation}
with
$$K^{''}(\hat{u})=\sqrt{x^2+t^2}.$$
Equation~\eqref{pxtexpdelta} provides for small $x/t$ the standard Gaussian behavior,
$P(x,t) \simeq \exp[ - x^2 / 2 t]$.
In the opposite limit  of  large $x$
\begin{equation}\label{eewwqwa19}
P(x,t) \simeq  \exp\left[ - x \ln\left( {2 x \over t}\right) \right],
\end{equation}
which demonstrates the exponential decay.
We may also write this solution like
\begin{equation}\label{eXPRateT}
P(x,t) \sim \frac{1}{\sqrt{2\pi K^{''}(\hat{u})}} \exp\left[ - t \mathcal{I}_t\left( {x \over t} \right) \right]
\end{equation}
with the rate function
\begin{equation}
\mathcal{I}_t(l) = 1+  l \sinh^{-1} (l)  - \sqrt{1 + l^2}.
\end{equation}
Notice that in Equation \eqref{eqes16} $P(x|n) = 0$  for $x> n$ since the particle walking on a lattice cannot reach a distance larger than the number of steps. Thus for standard random walks, with a fixed number of jumps, the exponential tails are not a generic feature. It implies that the fluctuation of the number of steps is crucial for the observation of exponential tails,  like those in Equation \eqref{eewwqwa19}.

\begin{figure}[htb]
 \centering
 \includegraphics[width=0.5\textwidth]{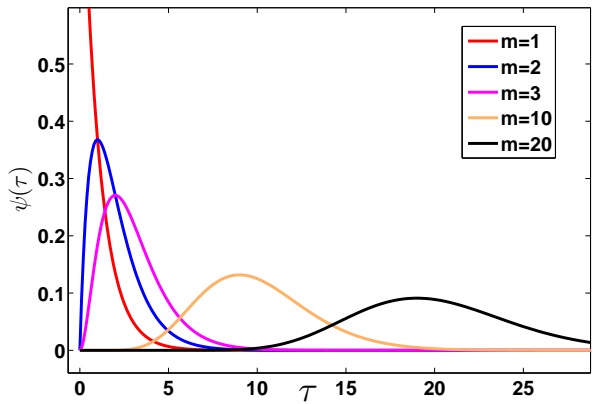}\\
 \caption{ Plot of Erlang PDF (Equation~\eqref{Erlang}) for various $m$. Increasing $m$ leads to decreasing  probability of obtaining small $\tau$ that leads to transition from bunching to anti-bunching (see discussion preceding Equation~\eqref{eqmandelparam}).
 %As shown in the figure, it's easy to detect small variables $\tau$ for a small $m$, showing bunching effects. While with the increase of $m$,  anti-bunching effects come into play (see below).
}\label{ErlangPDF}
\end{figure}

\section{The exponential tails for Erlang waiting time PDF}\label{sjdhhsiwq}

So far we exposed the exponential tails of $P(x,t)$ assuming
that the statistics of the numbers of jumps is Poissonian.
To advance the theory further we now consider $Q_t(n)$ for
the Erlang PDF of waiting times. For Erlang distribution, the  variance of the waiting time is finite. In the long time limit,  from the central limit theorem $n$ follows well known Gaussian distribution \cite{Cox1977Theory,Godreche2001Statistics}. This indicates that the spreading of the particles obeys Gaussian distribution for the central part of the distribution of the position. For more details see Appendix. \ref{APPENBeforea}.  For the tails, large number of jumps $n$ is
the cause for large displacement from the origin of the random
walker. Hence the limit $n$ large and $t$ fixed, or both these observables
large while their ratio is finite is of interest.

\subsection{The exponential tail of the number of events}

In this regard, two of us have found a general solution of the
problem~\cite{Barkai2020Packets}. The large $n$ limit of $Q_t(n)$ is controlled by the small $\tau$ limit of
the waiting time PDF, since to have many jumps within a finite
fixed time $t$,  the time intervals between
jumps must be made small. The main requirement we use here is that
$\psi(\tau)$ is analytical in the vicinity of $\tau\to 0$ and hence
\begin{equation}\label{ConDeqSTAS}
\psi(\tau) \sim C_A \tau^{ A}  + C_{A+1} \tau^{A+1}.
\end{equation}
Here $A$ is a non negative integer, and $C_A$ and $C_{A+1}$ are
coefficients. Then in the large $n$ limit, and $t$ fixed
we have \cite{Barkai2020Packets}
\begin{equation}\label{eqSTAS}
Q_t (n) \sim { \left\{ [ C_A \Gamma(1+A)]^{1 /(1+A)} t \right\}^{n(1+A)} \over \Gamma\left[ n(1+ A) + 1 \right] } \exp\left( t {C_{A+1}\over C_A}\right)
.
\end{equation}
Beyond $A$, $C_A$ and $C_{A+1}$ other properties of $\psi(\tau)$ are
irrelevant. Below, e.g. in Figure ~\ref{QtNtwoExp}, we call Equation \eqref{eqSTAS} the Qln formula  which denotes the limiting law of $Q_t(n)$ for $n\to\infty$.

We now consider an example, the Erlang distribution \cite{Forbes2011Statistical}
\begin{equation}\label{Erlang}
\psi(\tau) = { \tau^{m-1} e^{ - \tau} \over (m-1)!},
\end{equation}
 where $m$ is an integer and the mean $\langle \tau \rangle =m$, see Figure ~\ref{ErlangPDF}. Here exponential waiting times are recovered when $m=1$.
The Laplace transform of $\psi(\tau)$ is
\begin{equation}
\hat{\psi}(s)  =  {1 \over (1 + s)^m}.
\label{eqLAP}
\end{equation}
Thus, the Erlang PDF of order $m$ is the
$m$ fold convolution of the exponential PDF. As already mentioned, a well known formula Equation~\eqref{LaplaceQSN} yields the Laplace transform $t \to s$
 of $Q_t(n)$.
%
%\begin{equation}\label{LaplaceQSN}
%\hat{Q}_s (n) = {1 - \hat{\psi} (s) \over s} \hat{\psi}^n (s).
%\end{equation}
%
Inserting Equation (\ref{eqLAP})  and using the Laplace pairs
\begin{equation}
{1 \over s (1 + s)^{n m} } \leftrightarrow 1 - { \Gamma\left( nm , t \right) \over \Gamma(n m) },
\end{equation}
where $\Gamma(a,t)$ is the incomplete Gamma function \cite{Abramowitz1984Handbook}, we  find
\begin{equation}\label{exactElangQTN}
Q_t (n ) = { \Gamma\left( m n+m ,t  \right) \over \Gamma\left( n m + m\right)}
- {\Gamma \left( n m , t\right) \over \Gamma(n m ) },  \ \  ~~~ n\geq 1,
\end{equation}
and
\begin{equation}\label{exactElangQTNzero}
Q_t (0) = \Gamma(m,t) /\Gamma(m).
\end{equation}
Using the identity valid for a
positive integer $m$
\begin{equation}
\Gamma(m,y) = \left( m-1\right)! e^{-y} \sum_{j=0} ^{m-1} { y^j \over j!},
\end{equation}
we get
\begin{equation}
Q_t (n) = e^{-t} \sum_{j= nm} ^{nm + m-1} { t^j \over j!}, \ \ ~~~~~n \ge 0.
\label{eqEXPAN}
\end{equation}
We will soon use this expression to find an exact solution of the CTRW with Erlang PDF of waiting times.

\begin{figure}[htb]
 \centering
 \includegraphics[width=0.5\textwidth]{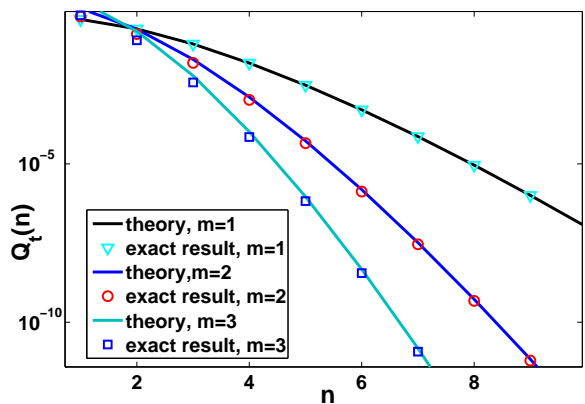}\\
 \caption{Probability to observe $n$ renewals when the waiting time is determined by Equation~\eqref{Erlang} for different $m$.   The solid lines are theoretical prediction Equation~\eqref{ElangExat} with $H(t,l)$ obtained from Equation~\eqref{aSYQNT} and the corresponding exact results, plotted by the symbols, are Equation~\eqref{exactElangQTN}. In our setting, we fix $\bar{t}=t/\langle\tau\rangle=t/m=1$ and change the value of $m$.
}\label{QntPDF}
\end{figure}

\begin{figure}[htb]
 \centering
 \includegraphics[width=0.5\textwidth]{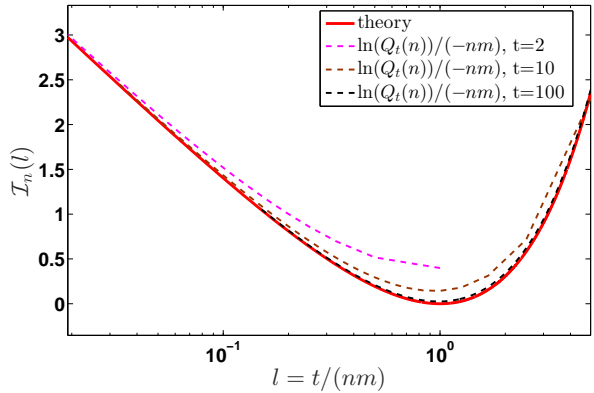}\\
 \caption{Rate function $\mathcal{I}_n(l)$ versus $l=t/(nm)$ for different observation times $t$ with $m=2$. The dashed lines are the plot of the function $\ln(Q_t(n))/(-nm)$ with $Q_t(n)$ calculated from Equation~\eqref{eqEXPAN}.
 Clearly, with the increase of the observation time $t$, the function $\ln(Q_n(t))/(-nm)$ versus $l$ approaches to the rate function $\mathcal{I}_n(l)$.
}\label{QntRateFuncitonVn}
\end{figure}

 From the above theorem, and the definition of the Erlang PDF we
have for this example $A=m-1$, $C_A= 1/(m-1)!$ and $C_{A + 1}= - C_A$.
Hence according to Equation \eqref{eqSTAS}, we have
\begin{equation}\label{elamngee101}
Q_t (n)  \sim e^{-t} { t^{nm } \over (nm)!}  .
\end{equation}
As expected this is the same as the  leading term in the expansion  Equation
\eqref{eqEXPAN}, when $t /n m\ll1$. We now find the rate function
of this example. Using the Stirling approximation $n!\sim\sqrt{2\pi n}(n/e)^n$, Equation~\eqref{eqEXPAN} reduces to
\begin{equation}\label{ElangExat}
Q_t (n)\sim{H(t,l)\over (2 \pi m n)^{1/2} }
\exp \left[ - m n \mathcal{I}_n(l) \right],
\end{equation}
where $\mathcal{I}_n(l) = -\ln(l) + l -1$, $l=t/nm$ and
\begin{equation}\label{exactHTL}
\begin{split}
  H(t,l)&=1+\frac{1}{\frac{1}{l}+\frac{1}{t}}+\frac{1}{(\frac{1}{l}+\frac{1}{t})(\frac{1}{l}+\frac{2}{t})}+\cdots+\frac{1}{(\frac{1}{l}+\frac{1}{t})\cdots(\frac{1}{l}+\frac{j-1}{t})}.
\end{split}
\end{equation}
For $l=t/nm\to0$, i.e., $n\to\infty$ and $m$ is fixed, the asymptotic behavior of Equation~\eqref{exactHTL} follows
\begin{equation}\label{aSYQNT}
\begin{split}
  H(t,l)&\sim1+l+l^2+l^{j-1}=\frac{1-l^{j-1}}{1-l}.
\end{split}
\end{equation}
This result is verified in Figure ~\ref{QntPDF}.
From Equation~\eqref{ElangExat}, the rate function becomes
\begin{equation}\label{rateNErlang}
\lim_{n \to \infty} { - \ln Q_t (n) \over m n} = \mathcal{I}_n (l),
\end{equation}
where the limit is valid when $l$ is fixed so here $t$ is made large. See the red solid line in  Figure  \ref{QntRateFuncitonVn}.
Finally we consider also
\begin{equation}
\lim_{t \to \infty} { - \ln Q_t (n) \over t} = \overline{l} \ln(\overline{l}) - \overline{l} + 1
\label{eqIQ}
\end{equation}
and here the limit is taken such that  $\overline{l}=  mn /t$ remains fixed.

{\bf Remark}
As mentioned we, use in our plots two graphical representations of the results, the distribution of observables of interest and  the rate
function. At least to the naked eye we see that in Figures (\ref{ExpGaussianPXT}, \ref{RateFunctionVersusT}, \ref{QntPDF}) the results converge already for relatively short time, e.g. t=2,  while in Figures
\ref{RateFunctionVersusX} and \ref{QntRateFuncitonVn}, we see that the convergence is achieved for much larger times, say $t=100$. This discrepancy dissolves if all the prefactors like $\ln\left(2K^{''}(\hat{u}) \right)$ are included, and not only the rate function.    The rate function formalism  is a limit where we take say $x\to\infty$ or $n\to \infty$. Taking the $log$ of the distribution, be it $P(x,t)$ or $Q_t (n)$ and dividing by a large number we get rid of the prefactors. Hence the two representations are of course not identical.

\subsection{The far tails of the positional PDF}
For a Gaussian PDF of jump lengths and Erlang's waiting
times we obtain a formal solution from Equations \eqref{PXT101} and \eqref{eqEXPAN}
\begin{equation}\label{ErlangPXTS102}
\begin{split}
P(x,t)=&\exp(-t)\sum_{n=1}^\infty\sum_{j=nm}^{nm+m-1}\frac{t^j}{j!}\frac{\exp(-\frac{x^2}{2n})}{\sqrt{2\pi n}}+\frac{\Gamma(m,t)}{\Gamma(m)}\delta(x).
\end{split}
\end{equation}
Similar to the previous examples, this function is easy to represent graphically for any reasonable $x$ and $t$ with a program like MATHEMATICA. Here our goal is to find the large $x$ behavior.

As briefly mentioned in the introduction, it was shown previously that $P(x,t)$ decays exponentially with the position $x$, more correctly like $|x|$ multiplied by a slowly varying function. We briefly outline the main result in \cite{Barkai2020Packets}. Using Cram\'{e}rs theorem \cite{Touchette2009large} it was shown that under certain conditions, in particular that the PDF of jump length decays faster than exponential decay,  $P(x|n)\sim \exp(-n(x/\delta n)^\beta)$ with $\beta>1$ and $|x|/n\to\infty$. Here we use Gaussian statistics for the jump  lengths so we have $\delta=\sqrt{2}$ and $\beta=2$. Then for large $x$ the main result in~\cite{Barkai2020Packets} reads
\begin{equation}\label{ejsdjjsl1}
P(x,t)\simeq \exp\left(-t\left\{\frac{|x|}{t}Z\left(\frac{|x|}{t}\right)-\frac{C_{A+1}}{C_A}\right\}\right),
\end{equation}
where
\begin{equation}
Z(y)=\frac{BW_0[g_1y^\beta]-g_0(A+1)}{W_0[g_1y^\beta]^{1/\beta}}
\end{equation}
with
\begin{equation}
B=\frac{g_0(A+1)}{\beta}+\frac{(g_0)^{1-\beta}}{\delta^\beta},
\end{equation}
$g_0=(\beta(\beta-1)/(A+1))^{1/\beta}/\delta$ and $g_1=[g_0(A+1)/(C_A\Gamma(A+1))^{1/(1+A)}]^\beta$.  Recall that the Lambert function $W_0(y)$ is a monotonically increasing function of $y$. Further $W_0(y)\sim \ln(y)$ for large $y$. Hence from Equation~\eqref{ejsdjjsl1} we learn that $P(x,t)$ decays  exponentially (neglecting log terms) as mentioned. This result establishes that the exponential decay of the PDF $P(x,t)$ is a universal feature of CTRW, in the same spirit as experimental and numerical evidence demonstrates in many examples.

\begin{figure}[htb]
 \centering
 \includegraphics[width=0.5\textwidth]{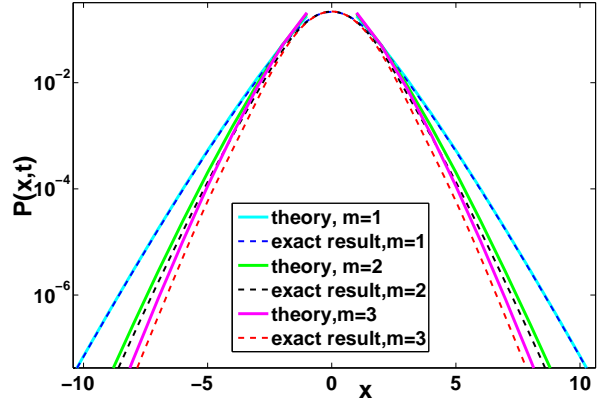}\\
 \caption{ Distribution of the position with Erlang distribution Equation~\eqref{Erlang}, where we chose  the rescaled time $\bar{t}=1$.
 Here the solid lines are the theory according to Equation~\eqref{PXTlang} and the corresponding exact result is obtained from Equation~\eqref{PXT101}. See Figure ~\ref{ErLangRatePXTTest} for the convergence of the theory.
}\label{ErlangPXTk5}
\end{figure}

To appreciate this result and further test it, let us derive it with our example (in this case the proof is somewhat easier as compared to the general case).  When $x\to\infty$ and $t$ fixed, say $t$ of order of $c\langle\tau\rangle$ and $c=1$ or $c=137$ etc, the terms contributing to the infinite sum
giving $P(x,t)$ are those with large $n$. Physically and
mathematically this is obvious, as to reach large $x$ in a finite  time we need many jumps since the average of displacement is finite. We therefore use large $n$ approximation of $Q_t(n)$ Equation~\eqref{eqEXPAN} and then
\begin{equation}
 P(x,t)\simeq \sum_{n=1}^\infty\frac{\exp(-t)t^{nm}}{(nm)!}\frac{\exp(-\frac{x^2}{2n})}{\sqrt{2\pi n}}.
\end{equation}
Only large $n$ terms contribute, hence including small $n$
in the summation is not a problem, since these terms are
of order $\exp(-x^2)$ while $|x|\to\infty$. We now switch to Fourier space, and find
\begin{equation}
\widetilde{P}(k,t)\simeq \exp(-t)S_m[t^m\exp(-k^2/2)]
\end{equation}
with the sum $S_m(y)=\sum_{n=0}^\infty y^n/(nm)!.$
The next step is to find the asymptotic behavior of $S_m(y)$ for large $y$. In order to do so, let us consider some special cases.
Hence for $m=1$, $S_1(y)=\exp(y)$, and similarly $S_2(y)=\cosh(\sqrt{y})\sim \exp(\sqrt{y})/2$,
\begin{equation}
\begin{split}
S_3(y)&=\frac{1}{3} e^{-\frac{\sqrt[3]{y}}{2}} \left(e^{\frac{3 \sqrt[3]{y}}{2}}+2 \cos \left(\frac{1}{2} \sqrt{3} \sqrt[3]{y}\right)\right)\\
&\sim \frac{1}{3}\exp(-y^{1/3}),
\end{split}
\end{equation}
etc. For a general $m$, we write the infinite sum as an integral
\begin{equation}
\begin{split}
S_m(y)&\to\int_0^\infty \frac{y^n}{(nm)!}dn\\
&=\frac{1}{m}\int_0^\infty \frac{(y^{1/m})^j}{(j)!}dj\\
&\sim \frac{1}{m}\exp(y^{1/m}).
\end{split}
\end{equation}
Below, we will use the above equation to calculate the far tails of the distribution of the position. As usual switching to $k=-iu$ we obtain the moment generating function, and taking the log, we get the cumulant generating function
\begin{equation}
K(u)\simeq -t+t\exp(u^2/(2m)).
\end{equation}
This is valid for large $u$ since that limit is the relevant
one for the calculations of $P(x,t)$ for large $x$, similar to
our previous examples. Here we neglected a $\ln(m)$ term
that is negligible in the limit under study. Using the
standard large deviation technique Equation~\eqref{Largedeviation101}, we find the solution
of $K^{'}(u)=x$ denoted $\hat{u}$ and this solves the equation
\begin{equation}
\hat{u}\exp\left(\frac{(\hat{u})^2}{2m}\right)=\frac{mx}{t},
\end{equation}
and hence we find yet again the Lambert type of solution
\begin{equation}\label{eqws1012}
\hat{u}=\sqrt{mW_0\left[m\left(\frac{x^2}{t^2}\right)\right]}
\end{equation}
for $x>0$ otherwise the right hand side of Equation~\eqref{eqws1012} has a negative sign.
If $m=1$, as expected, we relax to Equation~\eqref{ussss1}. Using the large deviation formula Equation~\eqref{Largedeviation101}, we find
\begin{equation}\label{PXTlang}
P(x,t)\sim \frac{\exp \left(-\sqrt{m} \left| x\right|\sqrt{W_0\left(\frac{m x^2}{t^2}\right)}+\frac{\sqrt{m} \left| x\right| }{\sqrt{W_0\left(\frac{m x^2}{t^2}\right)}}-t\right)}{ \sqrt{\left|\left| x\right|2 \pi m^{-1/2}\left(\sqrt{W_0\left(\frac{m x^2}{t^2}\right)} +\frac{1}{\sqrt{W_0\left(\frac{m x^2}{t^2}\right)}}\right)\right| }}.
\end{equation}
This  solution and the more general one Equation~\eqref{ejsdjjsl1}  are of course in agreement. As shown in Figure ~\ref{ErlangPXTk5}, the far tails of the distribution of the position follow  exponential decay  predicted by Equation~\eqref{PXTlang} since as mentioned $W_0(z) \sim \ln(z)$.
But as shown in Figure ~\ref{ErlangPXTk5}  the rate of convergence is slow, for example for $m=3$.

To check better the convergence issue, we consider the rate function.
Based on Equation~\eqref{PXTlang},
the rate function with respect to the position becomes
\begin{equation}\label{erlangrATEPXT}
\lim_{t\to\infty}-\frac{\ln(P(x,t))}{|x|}=\mathcal{I}_{x}(l)
\end{equation}
with $l=x/(t/\langle\tau\rangle)$ and
\begin{equation}
\mathcal{I}_{x}(l)=\sqrt{m}\left(W_0(l^2/m)-\frac{1}{W_0(l^2/m)}\right)+\frac{m}{|l|}.
\label{erlangrATEPXT02aa}
\end{equation}

\begin{figure}[htb]
 \centering
 \includegraphics[width=0.5\textwidth]{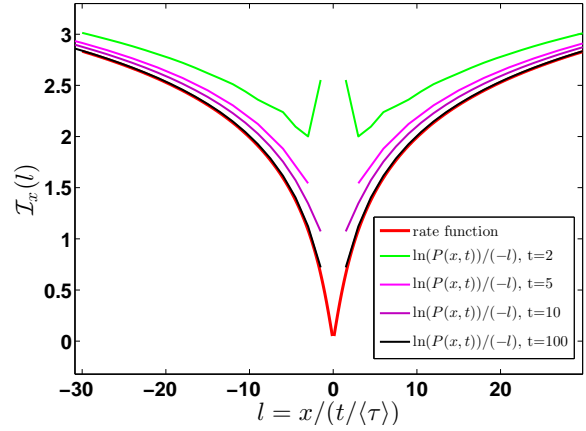}\\
 \caption{ Comparison of analytical prediction Equation~\eqref{erlangrATEPXT} (red line) for $\mathcal{I}_x(l)$ with  $\ln(P(x,t))/(-l)$. We show that  $\ln(P(x,t))/(-l)$ obtained from Equation~\eqref{ErlangPXTS102} converges to the rate function  Equation~\eqref{erlangrATEPXT} with the growing of the observation time $t$. Here we choose $m=3$.
}\label{ErLangRatePXTTest}
\end{figure}

It can be seen in Figure ~\ref{ErLangRatePXTTest} that with the increase of observation time $t$, the left-hand of Equation~\eqref{erlangrATEPXT} tends to  $\mathcal{I}_x(l)$ (Equation~\eqref{erlangrATEPXT02aa}) slowly. One contribution to this slow convergence effect, is that $t$ is not very large, the prefactor of $\exp(-|x|\mathcal{I}_x(l))$, i.e. $1/\sqrt{2\pi K^{''}(\hat{u})}$, is of importance; see Figure ~\ref{ErLangRatePXTTest}.

As expected, when $mx^2/t^2\to 0$, we get the Gaussian distribution
\begin{equation}
P(x,t)\simeq \exp\left(-\frac{x^2m}{2t}\right)
\end{equation}
and since $\langle \tau \rangle = m$ from Equation~\eqref{Erlang}  we have
$P(x,\bar{t}) \simeq \exp( - x^2/2 \bar{t})$
with $\bar{t} =t/\langle \tau \rangle$.
Recall that $W_0(y)\sim \ln(y)$ for $y\gg 1$, hence when $mx^2/t^2\gg 1$ we may approximate Equation~\eqref{PXTlang}  with
\begin{equation}
P(x,t)\simeq  \exp\left(-\sqrt{m} \left| x\right|  \sqrt{\ln\left(\frac{m x^2}{t^2}\right)}-t\right).
\end{equation}
Thus using the rescaled time $\bar{t}=t/\langle\tau\rangle$
we have
$$P(x,t)\simeq  \exp\left( -\sqrt{m} |x| \ln\left[ \frac{x^2}{ m\bar{t}^2} \right]-t\right),$$ implying that as we increase $m$ for fixed $\bar{t}$ we get  a suppression of the exponential tails.

\begin{figure}[htb]
 \centering
 \includegraphics[width=0.5\textwidth]{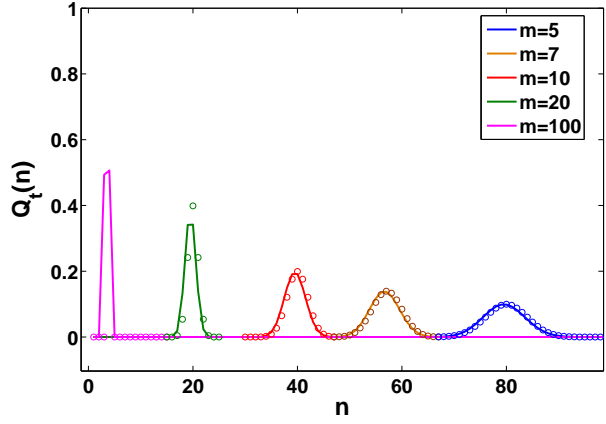}\\
 \caption{ $Q_t(n)$ for the case of Erlang PDF of waiting times, i.e. $\psi(\tau)$ is determined by Equation~\eqref{Erlang}. The solid lines describe Equation~\eqref{exactElangQTN} the exact behavior of $Q_t(n)$ for various $m$. Here we choose $t=400$.
 When $m$ increases the distribution narrows, indicating that fluctuations of $n$ disappear. We further plot the central part of $Q_t(n)$ using the symbols according to Gaussian approximation. Here for Gaussian approximation we use $Q_t(n)\sim \exp(-(n-t/m)^2/(2t/m^2))/\sqrt{2\pi t/m^2}
$ (see Appendix \ref{APPENBeforea}).
 %Distribution of $n$ for a finite time $t$, here $t=400$. The solid lines are the theoretical prediction Equation~\eqref{exactElangQTN} for different $m$. As the increase of $m$, $Q_t(n)$ becomes more narrow or restricted, indicting that fluctuations of $n$ disappear.
}\label{ErlangVsDelta}
\end{figure}

As we increase $m$ the condition on $x\gg \sqrt{m}\bar{t} $ implies that we may observe the nearly exponential decay of the packet  but only for a very large $x$, or only for very short times. This is because when we increase $m$, the waiting time exhibits a strong anti-bunching effect. Namely, to find a large $x$ we need to have a large fluctuations of $n$. For example if the average of $\langle n(t) \rangle\sim 2$ we may still see realizations with many jumps, leading to exponential decay in the far tails of $x$. However, this is less likely for large $m$ if compared with small $m$; {see Figure ~\ref{ErlangPDF}}. This is because when $m$ is large  we have vanishing fluctuations of $n$.  To quantify this we may use a tool from quantum optic, namely the Mandel $Q$ parameter \cite{Mandel1979Atomic} defined as
\begin{equation}
\label{eqmandelparam}
Q=\frac{\langle n^2\rangle-\langle n\rangle^2}{\langle n\rangle}-1.
\end{equation}
The case $Q<0$ is called sub-Poissonian (anti-bunching ) and $Q>0$ sup-Poissonian. If $Q=-1$, we have no fluctuations of $n$ at all.
For the Erlang PDF, and in the long time limit $Q=-1+1/m$. Namely, when $m\to\infty$ the fluctuations of $n$ vanish {
(see Figure  \ref{ErlangVsDelta})
}, and hence with it the effect of exponential tails of $P(x,t)$.
This is easy to see since in the absence of fluctuations of $n$, the number of jumps is fixed and since we have Gaussian jump lengths, the total displacement will be Gaussian and not exponential.  Thus we see that as we increase $m$, namely make the process more anti-bunched  we see less  exponential tails.
Anti-bunching means the effective repulsion of the dots on the time axis on which
jump events takes place. And this is controlled by the small $\tau$ behavior of $\psi(\tau)$. The anti-bunching $Q$ reduces
 to a negative value, and hence kills fluctuations of $n$ which are the key to the observation  of  exponential tails of $P(x,t)$.

\section{Bunching case of waiting time PDF}\label{sahshawa}
We further explore the behavior of probability to observe a large number of renewals $n$.  For some special distributions of the waiting times it is unwise simply to take the  $n\to \infty$ limit since  the rate of convergence to Equation~\eqref{eqSTAS} is very slow. Here we consider $\psi(\tau)$ as a  sum of two exponential waiting times PDFs
\begin{equation}\label{equsssmd101}
\psi(\tau)=\frac{1}{2}\left(\frac{1}{a}\exp\left(-\frac{\tau}{a}\right)+\frac{1}{b}\exp\left(-\frac{\tau}{b}\right)\right).
\end{equation}
We choose $a<b$ while the opposite situation is merely relabeling. The average  waiting time is $\langle\tau\rangle=(a+b)/2$. In what follows we shall fix  $\langle \tau \rangle=1$, so $a+b=2$ and this means that we have one free parameter say $a$ with $0<a<1$. We also consider the case where the jump size distribution is Gaussian with a variance equal unity and zero mean. This implies that in the long time limit, independently of the specific choice of $a$,  the mean square displacement
$\langle x^2\rangle\sim t$ and
  the process is Gaussian.
However, for short times
%and with respect to the exponential tails
the dynamics of $P(x,t)$ is of course $a$ sensitive. The question we wish to address: how does $a$ control the exponential tails of $P(x,t)$?

This is related to effect of bunching. Here for short waiting times we have $\psi(0)=(1/a+1/b)/2$. When $a=b=1$ we have a Poisson process, while when say $a\to 0$, we get a large value of $\psi(0)$. This means that there is an increased probability (compared to  Poissonian case) to obtain a jump shortly after a jump event. This is an effect of bunching where the jumps come in groups, and then separated by the relatively large waiting times. Such intermittent behavior is quantified with Mandel $Q$ parameter, which for large time is
\begin{equation}
Q=2(a-1)^2,
\end{equation}
where we assumed that $a+b=2$.
This is a super Poissonian behavior since $Q>0$. Note that in the previous example of the Erlang PDF, we had the opposite behavior, i.e. $Q<0$, since there $\lim_{\tau\to 0}\psi(\tau)=0$ for any $m\neq 1$.

\begin{figure}[htb]
 \centering
 \includegraphics[width=0.5\textwidth]{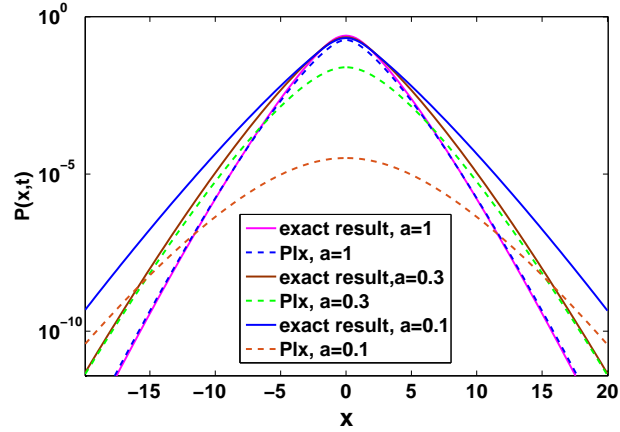}\\
 \caption{
$P(x,t)$ for the case where the jump lengths are Gaussian and $\psi(\tau)$ is given by Equation~\eqref{equsssmd101}. The solid line describe the exact behavior for various $a$, as provided by Equations~(\ref{PXT101},~\ref{twoExp1002}), while the corresponding large $|x|$ approximations (termed Plx) due to Equation~\eqref{ejsdjjsl1} described by the dashed lines. For small $a$, e.g., $a=0.1$, the convergence of the large $|x|$ approximation is slow and achieved only for very large values of $|x|$. See  Appendix \ref{APPENa} for a proper discussion of the  behavior in this limit.
 %The plot of the PDF of $x$ for different $a$.  The exact result and the Plx, describing the behavior of the distribution of the position for  $x\to\infty$, are shown by Equations~(\ref{PXT101},~\ref{twoExp1002})  and Equation~\eqref{ejsdjjsl1}, respectively.
% The figure shows that as we increase $a$, the nearly exponential decay of the tails is
%becoming faster. Thus as we decrease $a$,
%making the process more super-Poissonian,  the tails become easier to detect. Further, for a small $a$, here $a=0.1$ the theory Equation~\eqref{ejsdjjsl1} does not describe the exact solution, namely it works but only  for a very large $x$, so large making the theory non-practical as it captures values of $P(x,t)$ so small which are non-measurable. For this reason we develop a theory based on small $a$ limit  which is provided in Appendix \ref{APPENa}, see Figure ~\ref{PXTtwoExpSmalla} there.
}\label{PXTtwoExpDifferentA}
\end{figure}

\begin{figure}[htb]
 \centering
 \includegraphics[width=0.5\textwidth]{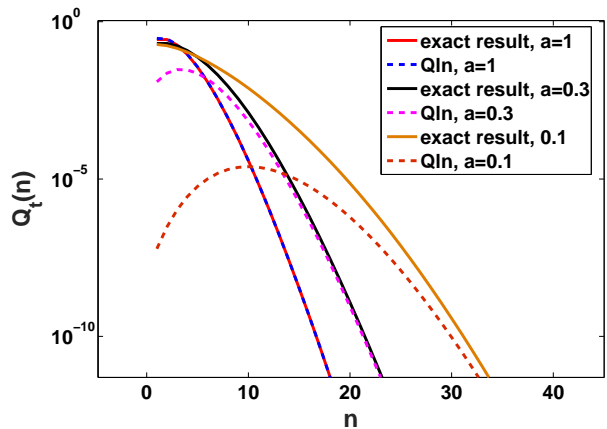}\\
 \caption{Probability $Q_t(n)$ of observing $n$ jump events versus $n$
while $\psi(\tau)$ is determined by Equation \eqref{equsssmd101}. The measurement time is $t=2$.
The solid lines describes the exact results for various $a$, i.e. Equation~\eqref{twoExp1002}, and the dashed lines are the corresponding Qln approximations, i.e. Equation~\eqref{eqSTAS}.
%We see that for small $a$, the Qln formula Equation~\eqref{eqSTAS} does not describe the exact result Equation~\eqref{twoExp1002} in this range. This simply means that the Qln formula works but for values of $n$ which are so large such that $Q_t(n)$ is really small (notice the small value of $Q_t(n)$).
%The solid lines arelculated fromEquation~\eqref{twoExp1002} and the corresponding dashed lines are the Qln expression  Equation~\eqref{eqSTAS}. For small $a$, we use the new theoretical prediction Equation~\eqref{sjejsjdj102}, shown by the symbols $`+'$, which exhibits exponential decay for not too large $n$ and agrees with the exact result. The related asymptotic behavior of $Q_t(n)$ is Equation~\eqref{sdjeskdje101}; see the dash-dotted line
%.
}\label{QtNtwoExp}
\end{figure}

We can obtain $Q_t(n)$ in terms of an integral; see details in Appendix \ref{APPENa}. Aside $Q_t(n)$  we also find $P(x,t)$, which is  compared to the previous theory when $x\to \infty$, i.e. Equation~\eqref{ejsdjjsl1}; see Figure  \ref{PXTtwoExpDifferentA}. Here we see that by   making $a$ smaller, namely making the process more bunched, we get a large exponential tail (see  Figure ~\ref{PXTtwoExpDifferentA}). Thus bunching makes the observation of the  exponential tails effect more readily achievable in experiments. In principle for any PDF of the waiting times (provided $\psi(\tau)$ is analytical in the vicinity of $\tau=0$)  an exponential decay of $P(x,t)$ is obtained, but this decay can be sometimes achieved for extremely large values of $x$.
We also observe that simply including the asymptotic behavior of $Q_t(n)$ in Equation~\eqref{eqSTAS} leads to slow convergence in the $a\to 0$ limit.
To further understand the effect of bunching  we also plot $Q_t(n)$ versus $n$; see Figure ~\ref{QtNtwoExp}. Strong bunching in this model, implies a relatively  high probability for seeing a large $n$ (many short time intervals between jumps since $a$ is  small). As we see, the decay of $Q_t(n)$ with $n$ is relatively slow when bunching is pronounced, and then the appearance of non negligible probability for large $n$, implies that particles can travel  large distances, and then the tails posses more statistical weight.  In Figure ~\ref{QtNtwoExp}  the slow convergence to asymptotic behavior of Equation~\eqref{eqSTAS} is observed in the $a\to 0$ limit and the delicate treatment of Appendix \ref{APPENa} is preferable.

%WANLI PLEASE CHECK THE MATH, FIND MY TYPOS.
%WE CAN ALSO PLOT ONE EXAMPLE TO SHECK CONVERGENCE. MAYBE THE
%FACT THTA RATE FUNCTIONS ARE $k$ INDEPENDENT IS NOCE TO DEMONSTRATE,
%SINCE WE MAY CHOOSE SAY $k=1,2, .. 5$ AND ALL RESULTS WILL FALL
%ON A MASTER CURVE SINCE I MEAN WHEN WE PLOT
%Equation (\ref{eqIQ}) the X AXIS IS $ k n /t$ the Y AXIS
%$Log[Q] / t $ AND $t$ is ACTUALLY LARGE (this is fine I guess since
%we have exact expressions, so we can plot for any $t$).

\section{conclusion}
Following the work of Kob and co-workers we have
formalized the problem of nearly exponential decay of
$P(x,t)$ using the CTRW framework. Exponential decay
is the rule, and it should be considered a natural
consequence of large deviation theory. In the long time
limit the packet of particles is typically Gaussian, hence
the phenomenon can be found/measured for intermediate
and short times. To describe the dynamics we have considered
a few examples where we may find exact solutions
to the problem. This allowed us to find the far tail, obtain
the rate function, and compare finite time solutions
with asymptotic expressions. We distinguished between
bunching and anti-bunching processes. These are determined
by the behavior of  $\psi(\tau)$ in the vicinity of of $\tau$ close to zero.
The short time behavior of  $\psi(\tau)$ determines the statistical
behavior of the number of jumps, when the latter
is large. And large number of jumps, leads the particle
to non-typical large $x$, where the phenomenon is found.
Using the example of  $\psi(\tau)$ expressed as a sum of two exponential
waiting times, we show that as we increase the
bunching effect, the exponential tail of $P(x,t)$ is more
pronounced. Similarly as we increase anti-bunching, by
increasing the parameter $m$ in the Erlang distribution, the
exponential tails are suppressed. Indeed as $m\to\infty$ we
get the usual random walk where $n$ is not
fluctuating and
then exponential tails cannot be found.

{These effects are
related to the rather universal behavior of $Q_t(n)$ found
for large $n$.
As we have demonstrated here, for large $n$ $Q_t(n)$ universally attains exponential decay (with log corrections).
This, as
mentioned, is valid for any  $\psi(\tau)$ which is analytic for
small $\tau$. Thus the agreement with the experimental observation
that finds exponential tails, is a manifestation
of the widely applicable CTRW model under study and
not merely consequence of a fitting  procedure.
This universal property of CTRW is the crucial difference from the diffusing diffusivity model~\cite{Leptos2009Dynamics,Chubynsky2014Diffusing,Jain2016Diffusion,Thomas2017Cytoplasmic,Cherstvy2019Non,Grebenkov2019unifying,Metzler2020Superstatistics,Hidalgo2020Hitchhiker}.
}

%\added{Should we use small $\tau$ in this paragraph?}
%the diffusing diffusivity model was used to uncover
%other stochastic properties of the process, like the first passage time distribution, and thus as a phenomenological
%tool it seems to us valuable.
%Though the latter popular
%method should be used with care, since our results
%while showing exponential decay in $x$ do not support a
%general scaling of the far tail with $x/\sqrt{t}$.

As mentioned in Sect. \ref{sjdhhsiwq}, in this manuscript we assume that the PDF of waiting time $\psi(\tau)$ is analytic at the vicinity of $\tau=0$. Clearly not all of the PDFs of waiting time satisfy this property, for example $\psi(\tau)=\tau^{-3/2}\exp[-1/(4\tau)]/2\sqrt{\pi}$. Roughly speaking, for this case the probability of performing many jumps at a finite time is much smaller than the cases studied here since $\psi(\tau\to 0)=0$. More specifically, for this example the analytical behavior  Equation \eqref{ConDeqSTAS} is not valid, and hence our main results do not apply.
This implies that the far tails of distribution of position decay faster than exponential tails.
%We expect that $P(x,t)\simeq \exp(-|x|^\beta/t^\eta)$ with $\beta>1$ and $\eta>0$.
It would be of interest to investigate  this problem, which is left for  future work.

Finally, let us mention a few other open problems:

\begin{itemize}
\item In the present case  we assumed that the jump process starts at time $t=0$. This is called an ordinary renewal process. If the processes started long before the measurement, we will have a modification of the PDF of the first waiting time \cite{Haus1987Diffusion,Hou2018Biased}. How does this effect the large $x$ behavior of $P(x,t)$? This issue seems important since the phenomenon can be found for relatively short times.

\item We focused on models that in the long time limit
converge to Gaussian statistics. What happens if  $\psi(\tau)$ is
fat tailed \cite{Bouchaud1990Anomalous,Metzler2000random} with diverging mean? Our results are certainly
valid for this case as well, however we did not explore
this in detail.

\item What happens in dimension $d>1$?

\item What are ideal waiting time PDFs and jump length
distributions, where exponential tails are pronounce and
if possible maintained for longer times. We showed how
this is related to bunching and anti-bunching, however
more refined work can help to clarify  better the widely
observed behaviour.

\item We used CTRW, instead one could use the noisy
CTRW model \cite{Jeon2013Noisy}. This adds to the jumps also noise when the
particle is waiting for its next jump. Thus noisy CTRW
is much more similar to real experiments.
\item Here we considered the decoupled CTRW, where
jump lengths and waiting times are uncorrelated. The
general framework of CTRW, goes beyond this simplification \cite{Klafter1994Levy,Marcin2017Aging}.
\item If the jump length PDF is sub-exponential, the far
tail of $P(x,t)$ will deviate from what we found here. Most
likely the principle of the single big jump \cite{Cistjakov1964theorem,Alessandro2019Single,Wang2019Transport} will hold in some
form, but the details of the theory are left unknown.

\item  Recently Dechant et al.  showed how the CTRW picture emerges from an under-damped Langevin description of a particle in a periodic potential \cite{Kindermann2017Nonergodic}. And then showed how this model can be used to analyse dynamics of Cesium atoms in optical lattices. Thus we expect to find also here exponential tails of packets, however  influence of  the control parameters of this phenomenon such as the depth of the optical potential,
the noise etc, are left unknown to us. Similarly, over damped Brownian motion in corrugated channels, a model of biophysical transport, is likely related to CTRW as a coarse grained description. In the former system exponential decay was already
explored in Ref. \cite{Li2019Non}. Thus exponential tails
are found both via Langevin dynamics and within CTRW,  the two approaches are related in some limits.
\end{itemize}

\section*{Acknowledgments}
E.B. acknowledges the Israel Science Foundations grant 1898/17.  S.B. was supported by the Pazy foundation grant 61139927.
W.W. was supported by Bar-Ilan University together with the Planning and Budgeting Committee fellowship program.

\appendix
\setcounter{figure}{0}
\renewcommand{\thefigure}{A\arabic{figure}}
\begin{appendices}

\begin{figure}
 \centering
 \includegraphics[width=0.5\textwidth]{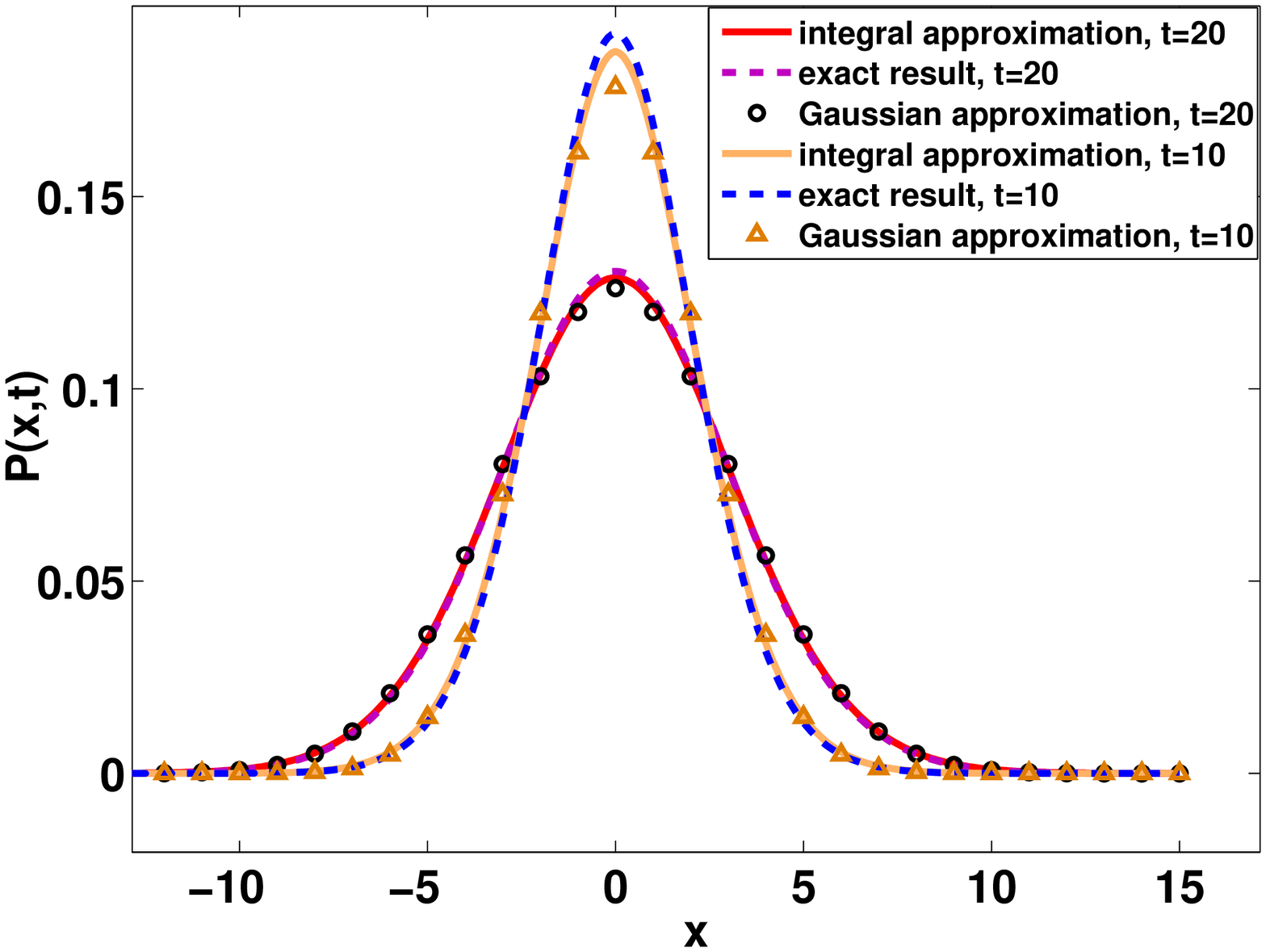}\\
 \caption{Comparison of exact result Equation~\eqref{PXT101} with integral approximation Equation~\eqref{APPENBeforea102} and Gaussian approximation Equation~\eqref{APPENBeforea103}. Here we choose $m=2$ and the observation time  $t=10$ and $20$.
}\label{GaussiainInt}
\end{figure}
\section{PDFs of $n$ and $x$ in the long time limit}\label{APPENBeforea}
When the waiting time has a finite variance and the observation time $t$ is large, $Q_t(n)$ obeys the limiting law described by Gaussian distribution \cite{Cox1977Theory,Godreche2001Statistics}
\begin{equation}\label{APPENBeforea101}
Q_t(n)\sim \frac{\exp \left(-\frac{\left(n-\langle n(t)\rangle\right)^2}{2(\sigma_n)^2}\right)}{\sqrt{2\pi (\sigma_n)^2}}.
\end{equation}
In turn $\langle n(t)\rangle$ and the variance of $n$, i.e., $\langle n^2(t)\rangle-\langle n(t)\rangle^2$ denoted as $(\sigma_n)^2$, can be obtained from
\begin{eqnarray*}
\langle n(t)\rangle&=&\mathcal{L}^{-1}\left[\frac{\hat{\psi}(s)}{s(1-\hat{\psi}(s))}\right]\\
\langle n^2(t)\rangle&=&\mathcal{L}^{-1}\left[\frac{\hat{\psi}(s)(1+\hat{\psi}(s))}{s(1-\hat{\psi}(s))^2}\right],
\end{eqnarray*}
where $\mathcal{L}^{-1}[\hat{g}(s)]$ means the inverse Laplace transform of $\hat{g}(s)$. In the limit of long $t$, we have $\langle n(t)\rangle\sim t/\langle\tau\rangle$ and $(\sigma_n)^2\sim t(\langle\tau^2\rangle-\langle\tau\rangle^2)/\langle\tau\rangle^3$.

We now consider the model with Erlang waiting time Equation~\eqref{Erlang} in the main text and Gaussian jump length $f(\chi)=\exp(-\chi^2/2)/\sqrt{2\pi}$.
In the long time limit we have $\langle n(t)\rangle \sim t/m$ and $(\sigma_n)^2\sim t/m^2$.
According to  Equation~\eqref{PXT101}, for large $t$, $P(x,t)$ reads
\begin{equation}\label{APPENBeforea102}
P(x,t)\simeq\int_0^\infty \frac{\exp(-x^2/(2n))}{\sqrt{2\pi n}}\frac{\exp \left(-\frac{\left(n-\langle n(t)\rangle\right)^2}{2(\sigma_n)^2}\right)}{\sqrt{2\pi (\sigma_n)^2}}dn.
\end{equation}
Figure \ref{GaussiainInt} demonstrates that Equation~\eqref{APPENBeforea102}, plotted by the solid lines, is an excellent approximation.
In the limit $t\to \infty$,   Gaussian distribution is found
\begin{equation}\label{APPENBeforea103}
P(x,t)\sim \frac{1}{\sqrt{2 \pi t/m}}\exp \left(-\frac{x^2}{2t/m}\right),
\end{equation}
where we used the relation $Q_t(n)\sim \delta(n-t/m)$.  Here we stress that Equation~\eqref{APPENBeforea103} is just valid for the central part of the distribution of the position, while the large deviation theory developed in the main text is applicable  in the large x limit.

\section{Calculation of $Q_t(n)$ and $P(x,t)$ with waiting time following Eq.~(\ref{equsssmd101})}\label{APPENa}

\begin{figure}
 \centering
 \includegraphics[width=0.5\textwidth]{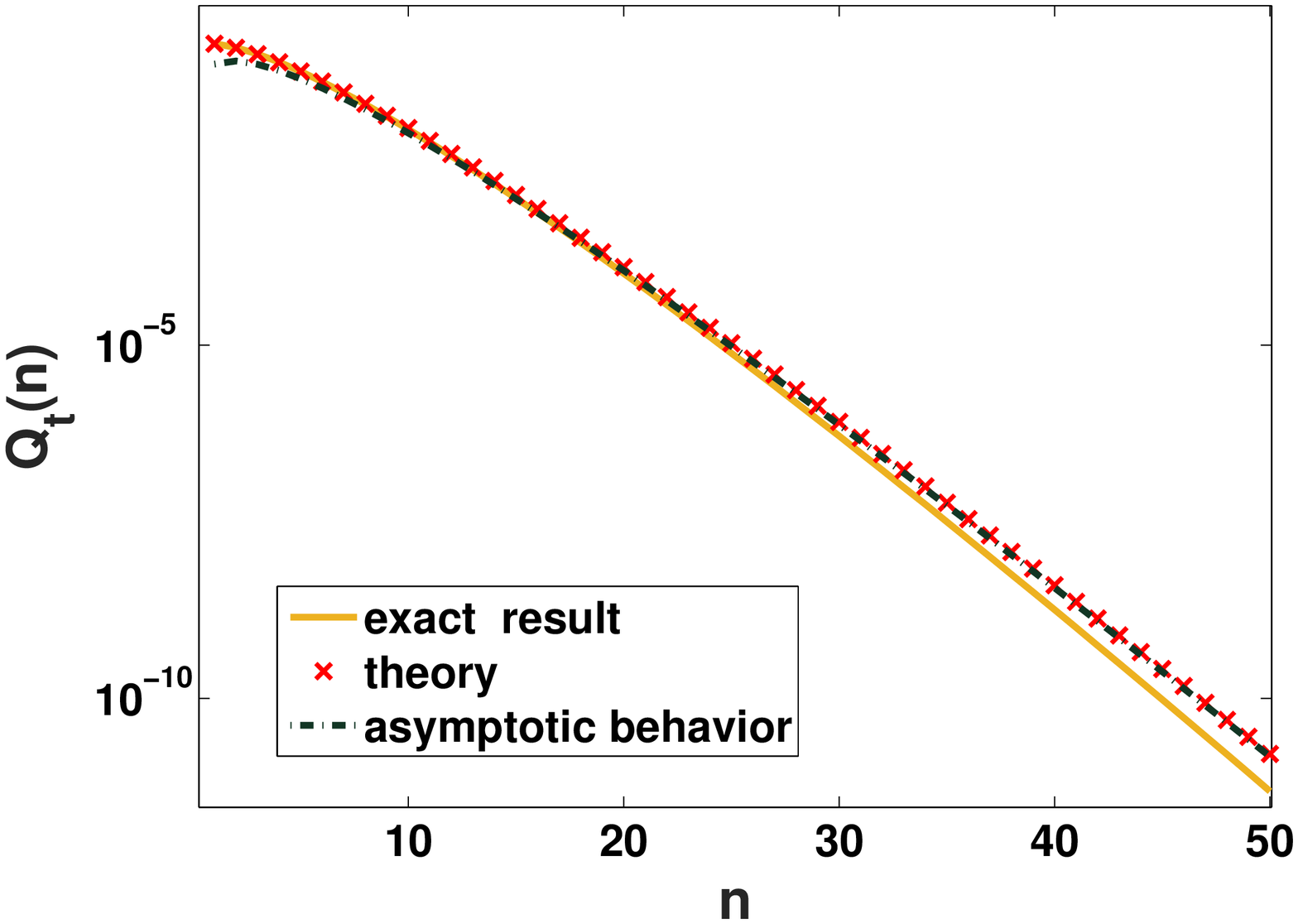}\\
 \caption{
 Probability $Q_t(n)$ of observing $n$ jump events versus $n$
while $\psi(\tau)$ is determined by Equation \eqref{equsssmd101}, for the case of $a=0.01$ while the measurement time is $t=2$. The solid line describes the exact result, according to Equation~\eqref{twoExp1002}. The  approximation in the limit of $a\to0$ (Equation~\eqref{sjejsjdj102}) is described by symbols and the dashed line is the asymptotic behavior of this approximation  (Equation~\eqref{sdjeskdje101}).
 }\label{QtNtwoExpNew}
\end{figure}

Here the aim is to find the exponential tails of the distribution of the number of renewals and the position.
As mentioned in the main text, we fix the mean of waiting times, namely $\langle\tau\rangle=(a+b)/2=1$. From Equation~\eqref{LaplaceQSN}, we have
\begin{equation}\label{twoExp1001}
\hat{Q}_t(s)=\frac{(\frac{1}{as+1}+\frac{1}{bs+1})^n}{s2^n}-\frac{(\frac{1}{as+1}+\frac{1}{bs+1})^{n+1}}{s2^{n+1}}.
\end{equation}
Thus, in real space, the formal solution is
\begin{equation}\label{twoExp1002}
Q_t(n)=\sum_{j=0}^nG(t,n,j)+\sum_{j=0}^{n+1}G(t,n+1,j)
\end{equation}
with
\begin{equation}
\begin{split}
G(t,n,j)&=\frac{n!}{2^nj!(n-j)!} \int_0^t\left(1-\frac{\Gamma(j,\frac{t-y}{a})}{\Gamma(j)}\right)\frac{\exp(-\frac{y}{b})(\frac{y}{b})^{n-j-1}}{b\Gamma(n-j)}dy.
\end{split}
\end{equation}

Let $a=0.01$. So $C_{A}=(1/a+1/b)/2\sim 1/(2a)$ and $C_{A+1}\sim-1/(2a^2)$ according to Equation~\eqref{ConDeqSTAS}. This indicates that $\exp( tC_{A+1}/C_A)\propto \exp(-t/a)=\exp(-200)$. It means that the tails are sensitive to value $a$ and it's not measurable in real experiment for such a small $a$. As shown in Figure ~\ref{QtNtwoExp}, for small $a$, the  limiting law Equation~\eqref{eqSTAS} loses its role in real experiment since it is too small to observe this phenomenon.
Below, the aim is to find a new formula to predicte such behavior. In the limit of $a\to 0$, Equation~\eqref{twoExp1001} reduces to
\begin{equation}\label{sjejsjdj101}
Q_t(s)\sim \frac{(s+1)^n}{(2s+1)^{n+1}}.
\end{equation}
One can do the similar calculation, like Equation~\eqref{twoExp1001}, but there is a simple way. From Cauchy's integral formula, the inverse Laplace transform of Equation~\eqref{sjejsjdj101}  yields
\begin{equation}\label{sjejsjdj102}
Q_t(n)\sim \frac{1}{2^{n+1}n!}\frac{d^n}{d s^n}[\exp(st)(s+1)^n]|_{s=-\frac{1}{2}},
\end{equation}
where we used the fact that Equation~\eqref{sjejsjdj101} has poles at $s=-1/2$. The equation \eqref{sjejsjdj102}, plotted by the symbols $(`+')$ in Figure ~\ref{QtNtwoExpNew}, is consistent with the exact result for $a=0.01$.
For this case, the law Equation~\eqref{eqSTAS} should be modified since the rate of convergence is extremely slow (for example, see the solid line for $a=0.1$ {in Figure ~\ref{QtNtwoExp}}). Another interesting problem is how the far tail of the distribution of  $n$ decays with respect to $n$.
Note that the limit between $a$ and $n$ is a bit subtle. Let $a$ be small and $n$ be large instead of infinity.
Using the saddle point approximation, we get
\begin{equation}
Q_t(n)\sim\frac{\exp \left(\frac{1}{4} \left(\sqrt{8 n t+t^2+4 t+4}-3 t+2\right)-(n+1) \ln \left(\frac{\sqrt{8 n t+t^2+4 t+4}-t+2}{2 t}\right)+n\ln \left(\frac{\sqrt{8 n t+t^2+4 t+4}+t+2}{4 t}\right)\right)}{\sqrt{32 \pi  \left| t^2 \left(\frac{n+1}{\left(-t+\sqrt{t^2+8 n t+4 t+4}+2\right)^2}-\frac{n}{\left(t+\sqrt{t^2+8 n t+4 t+4}+2\right)^2}\right)\right| }}.
\end{equation}
The corresponding asymptotic behavior of large $n$ follows
\begin{equation}\label{sdjeskdje101}
Q_t(n)\sim \frac{\exp \left(-\ln \left(\frac{\sqrt{8 n t}}{2 t}\right)+\sqrt{2 n t}-n \ln (2)+\frac{1}{2}-\frac{3 t}{4}\right)}{\sqrt{4 \pi  t^2 \left| \frac{\sqrt{2}}{\sqrt{t n}}\right| }},
\end{equation}
which is plotted by the dash-dotted line in Figure ~\ref{QtNtwoExpNew}.
In this limit, we still find the universal exponential decay of the far tail since  the leading term of Equation~\eqref{sdjeskdje101} is $n\ln(2)$.

%%%%%%%%%%%%%%%%%%%%%%%%%
\begin{figure}
 \centering
 \includegraphics[width=0.5\textwidth]{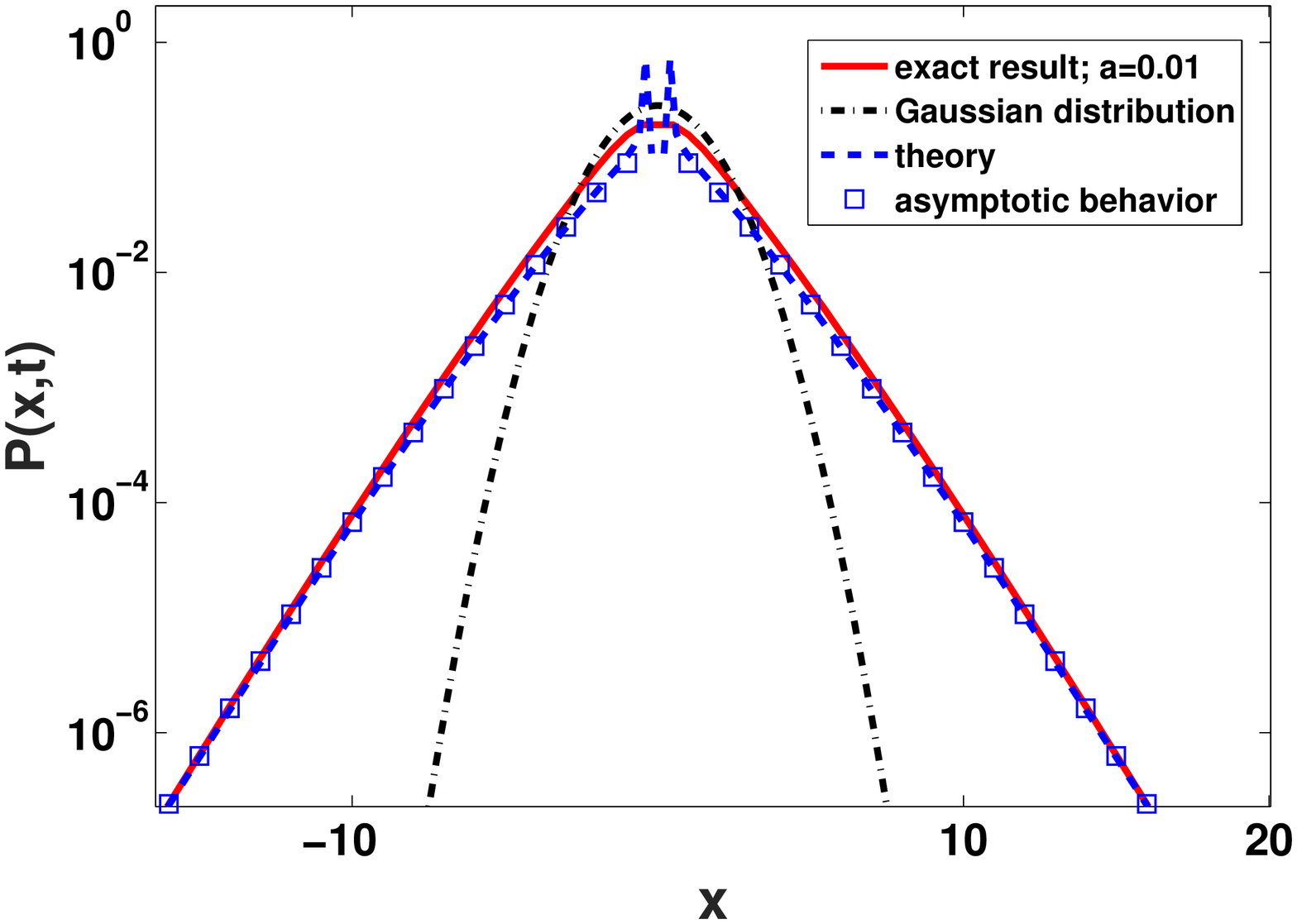}\\
 \caption{$P(x,t)$ for the case where the jump lengths are Gaussian and $\psi(\tau)$ is given by Equation~\eqref{equsssmd101}, in the limit $a\to 0$. The exact result (Equations~(\ref{PXT101},~\ref{twoExp1002})) for the case of $a=0.01$ is described by the solid line. The theoretical approximation in the limit $a\to 0$ (Equation~\eqref{ssdjflsldk101}) is presented by the dashed line and and the symbols  present the  asymptotic behavior of the approximation (Equation~\eqref{ssdjflsldk102}). The Gaussian distribution (dashed-dotted line) is provided for comparison.
 %is descri Here the exact result, shown by Equation~\eqref{PXT101}, is obtained from Equation~\eqref{twoExp1002} and $P(x|n)=\exp(-x^2/(2n))/\sqrt{2n}$. Our new theory is the dashed line calculated from Equation~\eqref{ssdjflsldk101}, whose asymptotic behavior is predicted by Equation~\eqref{ssdjflsldk102} plotted by symbols `$\Box$'. In our setting, $t=2$ and $a=0.01$.
}\label{PXTtwoExpSmalla}
\end{figure}

%%%%%%%%%%%%%%%%%%%%%%%

Utilizing Equation~\eqref{sdjeskdje101} and $P(x|n)=\exp(-x^2/(2n))/\sqrt{2n}$, and using the saddle point method again, we find
\begin{equation}\label{ssdjflsldk101}
P(x,t)\sim \frac{\exp \left(\sqrt{t} \sqrt[4]{\frac{2}{\ln (2)}} \sqrt{\left| x\right| }-\ln \left(\frac{\sqrt[4]{\frac{2}{\ln (2)}} \sqrt{\left|x\right| }}{\sqrt{t}}\right)-\frac{1}{4} \ln \left(\frac{\left|x\right| }{\sqrt{2 \ln (2)}}\right)-\sqrt{2\ln (2)} \left| x\right|-\frac{3t}{4}+\frac{1}{2}\right)}{\sqrt{4\pi\sqrt{2}t^{3/2}} \sqrt{\left|-\frac{2\sqrt{2}x^2 \ln^{\frac{3}{2}}(2)}{\left| x\right|^3}-\frac{\sqrt{t} \ln ^{\frac{3}{4}}(2)}{2^{3/4} \left| x\right| ^{3/2}}+\frac{3 \ln (2)}{2 \left| x\right| ^2}\right| }},
\end{equation}
For large $x$, we obtain
\begin{equation}\label{ssdjflsldk102}
P(x,t)\sim \frac{\exp \left( \sqrt[4]{\frac{2}{\ln (2)}} \sqrt{t\left| x\right| }+\ln \left(\frac{\sqrt{t \ln (2)}}{\sqrt[4]{\left| x\right| }}\right)-\sqrt{2\ln (2)} \left| x\right| -\frac{3 t}{4}+\frac{1}{2}\right)}{\sqrt{24 \pi  t^{3/2} \ln(2)}}.
\end{equation}
This is verified by the symbols in Figure ~\ref{PXTtwoExpSmalla} with $a=0.01$.
Clearly, the leading term $\sqrt{2\ln(2)}|x|$ in the numerator of the above equation is responsible for exponential decay. While, note that with the growing of $x$ and $n$,  Equation~\eqref{ssdjflsldk101} fails since for large $x$ the parameter $a$ comes into play, see  the symbols in Figure ~\ref{QtNtwoExpNew}. On the contrary, the law Equation~\eqref{eqSTAS} will work for the extremely large $n$, which is difficult to see for small $a$ if compared with Equation~\eqref{ssdjflsldk101}. As mentioned in the main text this is related to bunching effects.
\end{appendices}


\begin{thebibliography}{10}
\bibitem{Eli2020SRW}
For standard random walks, namely the case where the number of jumps $n$ is
  fixed, and when the pdf of jump length is exponential, one can show that
  ${P}(x|n)$ has exponential tails for any $n$. in the manuscript, we wish to
  avoid this nearly trivial case, namely we want to show that starting with say
  a gaussian pdf of jump lengths, we get an exponential tail for ${P}(x,t)$.we
  need the effect of bunching, many jumps close one to another, then maybe some
  inactive phase.
\end{thebibliography}
\end{document}